\documentclass[aps,twocolumn,floats,prl,nofootinbib]{revtex4}
\usepackage{graphics,graphicx,epsfig}
\usepackage{amssymb,color}
\usepackage{amsmath}
\usepackage{epsf,epstopdf,wrapfig}

\begin{document}

\title{Long time scales, individual differences, and scale invariance  in animal behavior}

\author{William Bialek$^{1,2}$ and Joshua W. Shaevitz$^1$}
\affiliation{$^1$Joseph Henry Laboratories of Physics and Lewis--Sigler Institute for Integrative Genomics, Princeton University, Princeton NJ 08544}
\affiliation{$^2$Center for Studies in Physics and Biology, Rockefeller University, New York NY 10065}

\begin{abstract}
The explosion of data on animal behavior in more natural contexts highlights the fact that these behaviors  exhibit correlations across many time scales.   But there are major challenges in analyzing these data:  records of behavior in single animals have fewer independent samples than one might expect; in pooling data from multiple animals, individual differences can mimic long--ranged temporal correlations; conversely long--ranged correlations can lead to an over--estimate of individual differences.  We suggest an analysis scheme that addresses these problems directly,  apply this approach to data on the spontaneous behavior of walking flies, and find evidence for scale invariant correlations over nearly three decades in time, from seconds to one hour. Three different measures of correlation are consistent with a single underlying scaling field of dimension $\Delta = 0.180\pm 0.005$.
\end{abstract}

\date{\today}

\maketitle

Animals, including humans, exhibit behaviors with structure on many time scales.  In one view, the many time scales result from distinct processes, perhaps organized hierarchically \cite{dawkins_76,simon_82}.  In another view, the wide range of time scales emerges from interactions among many underlying degrees of freedom, perhaps approaching a nearly scale invariant continuum \cite{bak_13}.  Scale invariance is especially tantalizing because of possible links to the renormalization group and critical phenomena \cite{wilson_79,cardy_96,sethna_06}.  

The literature on scale invariance in living systems is dominated by theoretical arguments:  why it might be advantageous for organisms to operate in this regime, or how apparent signatures of criticality and scale invariance might have  more prosaic explanations. This is an opportune moment to revisit the experiments because we have seen an explosion of quantitative data on animal behavior under  more naturalistic conditions.  Examples include  the variability of eye movement trajectories in primates \cite{osborne+al_05};  the postural dynamics of freely moving {\em C elegans} \cite{stephens+al_08,ahamed+al_21},  walking flies \cite{branson+al_09,berman+al_14,berman+al_16}, and mice \cite{wiltschko+al_15,marshall+al_20};  behavioral bouts in zebrafish larvae \cite{reddy+al_20,johnson+al_20, ghosh+rihel_20}; birdsong and other acoustic sequences \cite{markowitz+al_13,kershenbaum+al_14,kershenbaum+al_16}.  
High resolution video imaging and efficient AI tools  combine  to make these approaches more generally applicable \cite{mathis+al_18,pereira+al_19,pereira+al_22}, and these data have brought renewed attention to the wide range of time scales in  behavior   \cite{brown+bivort_17,berman_18,datta+al_19,pereira+al_20,mathis+mathis_20,bialek_20,alba+al_20}.  Although we focus here on the behavior of individual organisms, there is strong evidence for scale invariance in collective behaviors of flocks and swarms \cite{cavagna+al_18}.

We can characterize a system either by measuring the mean behavior in response to external perturbations or by analyzing the correlations in spontaneous fluctuations of behavior.  In thermal equilibrium these are equivalent \cite{sethna_06}, and there are ideas about how this connection can be generalized to more complex systems.  Here we focus on the analysis of correlations in the  spontaneous  behaviors of individual organisms, and we want to do this over a wide range of time scales. But if correlations are sufficiently long--ranged, then  in a single  behavioral trajectory we never have truly independent samples, and many statistical intuitions break down.  Also, if we are interested in the longest time scales that we can access in a given experiment, then by definition we don't have many samples, independent or not.  In order to increase statistical power, quantitative studies of animal behavior often average over multiple individual organisms, but this makes sense only if the different organisms behave in the same way.   In trying to measure correlations over long times,  individual differences are an  important confounding factor, essentially because each individual has an infinite memory of its own identity. 

The goal of this work is to disentangle the non--independence of samples, individual differences, and genuinely long--ranged correlations.  Although the issues are general, we ground our discussion in the analysis of experiments on the behavior of walking flies \cite{berman+al_14,berman+al_16}.  We will see that these data provide evidence for scale invariant correlations over nearly three decades, from time scales of seconds to one hour.

The raw data are high resolution video of fruit flies, from an inbred laboratory stock, walking in a featureless arena.   Video frames are of duration $\Delta t = 0.01\,{\rm s}$, single trajectories are of length $T_{\rm max} = 3600\,{\rm s}$, and we count time $t = 1,\, 2,\, \cdots$   in units of $\Delta t$; there are $N_f = 59$ individual flies in the data set \cite{berman+al_14,berman+al_16}.  Through a combination of linear and nonlinear dimensionality reduction these data can be embedded in a low dimensional space.  In this space the fly repeatedly visits small neighborhoods and then jumps to another,   defining $122$ discrete behavioral states, which also can be seen as peaks in the probability distribution over the continuous space; in addition there is a null state.  Some of these states have names (e.g., different forms of grooming) and some do not.  The same states can be identified across related species of flies, and  the distribution over these states  varies systematically with evolutionary distance \cite{hernandez+al_21}.  Trajectories through the discrete state space are strongly non--Markovian \cite{berman+al_16}, and compressed versions of these state sequences are described by models with nearly scale invariant interactions \cite{alba+al_20}.  

We  define $n_{\rm i}  (t) = 1$ if an individual fly is in state $\rm i$ in the small bin of duration $\Delta t$ surrounding time $t$.  We can characterize the correlations, summed over the individual states, by the (connected) 2--point function
\begin{eqnarray}
C(t,t') &=& \sum_{{\rm i}=1}^{N_s} \left[  \langle n_{\rm i}(t) n_{\rm i} (t') - \langle n_{\rm i}\rangle^2 \right] 
\label{defC}\\
&=& P_c(t,t') - P_c({\rm ind}) ,
\end{eqnarray}
where $P_c (t,t') $ is the coincidence probability of finding a fly in the same state at times $t$ and $t'$, and 
$P_c ({\rm ind})$ is the coincidence probability if we draw two independent samples from the distribution over states.  If the behavior is statistically stationary then $C(t,t') = C(t-t')$.  

Perhaps surprisingly, one serious problem in estimating these correlations from data is subtracting the mean or disconnected part. If we estimate $\langle n_{\rm i}\rangle$ as a time average over the behavioral trajectory of a single fly, then the correlation function must integrate to zero over the finite duration of our observations \cite{intC0}.  If trajectories are very long compared with the time scales of correlation this doesn't matter, but long--ranged correlations will be significantly distorted.  An alternative is to estimate $\langle n_{\rm i}\rangle$ as an average both over time and over an ensemble of flies.  But if individuals  have even slightly different mean behaviors this also distorts the correlation function, even violating the condition that the connected correlation $C(\tau )$ should vanish at large $|\tau|$. 

To disentangle long--ranged correlations and individual differences, we take a direct approach.  We estimate the probability that fly $\alpha$ occupies state $\rm i$ by averaging over a window of duration $T$,
\begin{equation}
\hat P_{\rm i}^\alpha (T) = {{\Delta t}\over T}\sum_{t=1}^{T/\Delta t} n_{\rm i}^\alpha (t) .
\end{equation}
With an ensemble of $N_f$ flies, we can then estimate the variance across individuals, summed over states:
\begin{equation}
\Phi_2(T)   = \sum_{{\rm i}=1}^{N_s} {1\over {N_f}} \sum_{\alpha = 1}^{N_f} \left[ \hat P_{\rm i}^\alpha (T) - {1\over {N_f}} \sum_{\beta = 1}^{N_f} \hat P_{\rm i}^\beta (T) \right]^2 .
\end{equation}

If there are no individual differences, then the only reason we see any variance across the ensemble of flies is because of statistical errors, that is because our estimates $\hat P_{\rm i}^\alpha (T)$ are based on a finite number of samples and thus differ from the true probabilities $P_{\rm i}^\alpha$.  In this case $\Phi_2(T)$ should get smaller at larger $T$ and eventually vanish.  But if there are true individual differences, then $\Phi_2(T\rightarrow\infty )$ will measure the variance of these differences.  Formally,
 \begin{equation}
\langle \Phi_2(T) \rangle = \Phi_{2c}(T)+ \Phi_{2,{\rm ind}}.
\end{equation}
Here the first term comes from the (connected) correlations in the behaviors of individuals in Eq (\ref{defC}),
\begin{equation}
\Phi_{2c}(T) = {1\over {N_f}}  \sum_{\alpha =1}^{N_f} \left({{\Delta t}\over T} \right)^2\sum_{t = 1}^T \sum_{t' = 1}^T C^\alpha (t - t') ,
\label{phi_c}
\end{equation}
where we assume stationarity, while the second term comes from individual differences
\begin{equation}
\Phi_{2,{\rm ind}} = {1\over {N_f}} \sum_{\alpha =1}^{N_f} \sum_{{\rm i}=1}^{N_s} \left[ P_{\rm i}^\alpha  -   {1\over {N_f}} \sum_{\beta =1}^{N_f}P_{\rm i}^\beta \right]^2 .
\end{equation}

If correlations in the behavioral trajectories of individuals are short--ranged, then at value of $T$ larger than the correlation time we will see the connected part $\Phi_{2c}(T) \sim 1/T$.  This corresponds to the intuition that variances should decay as the inverse of the number of independent samples.  On the other hand if correlations are long--ranged, with a power--law decay, 
$C^\alpha (\tau \rightarrow \infty ) \sim {1/ {|\tau|^{2\Delta}}}$, then $\Phi_{2c}(T) \sim 1/T^{2\Delta}$ at large $T$.  Power--law decays are referred to colloquially as scale invariant because there is no characteristic time over which the correlations decay.  But scale invariance is more than the observation of a single power law.  If behavior is determined by an internal variable that undergoes scale invariant fluctuations then we should see related power laws in different moments of these fluctuations \cite{cardy_96}.  Concretely, we define
\begin{equation}
\Phi_n(T)   = \sum_{{\rm i}=1}^{N_s} {1\over {N_f}} \sum_{\alpha = 1}^{N_f} \left[ \hat P_{\rm i}^\alpha (T) - {1\over {N_f}} \sum_{\beta = 1}^{N_f} \hat P_{\rm i}^\beta (T) \right]^n ,
\end{equation}
and the prediction of scale invariance is that
\begin{equation}
\langle \Phi_n(T\rightarrow\infty )\rangle  = A_n\left({{\Delta t}\over T}\right)^{\gamma_n} +B_n ,
\label{fit}
\end{equation}
with $\gamma_n = n\Delta$, and $\Delta$ is the ``scaling dimension.''   

\begin{figure*}[t]
\includegraphics[width = \linewidth]{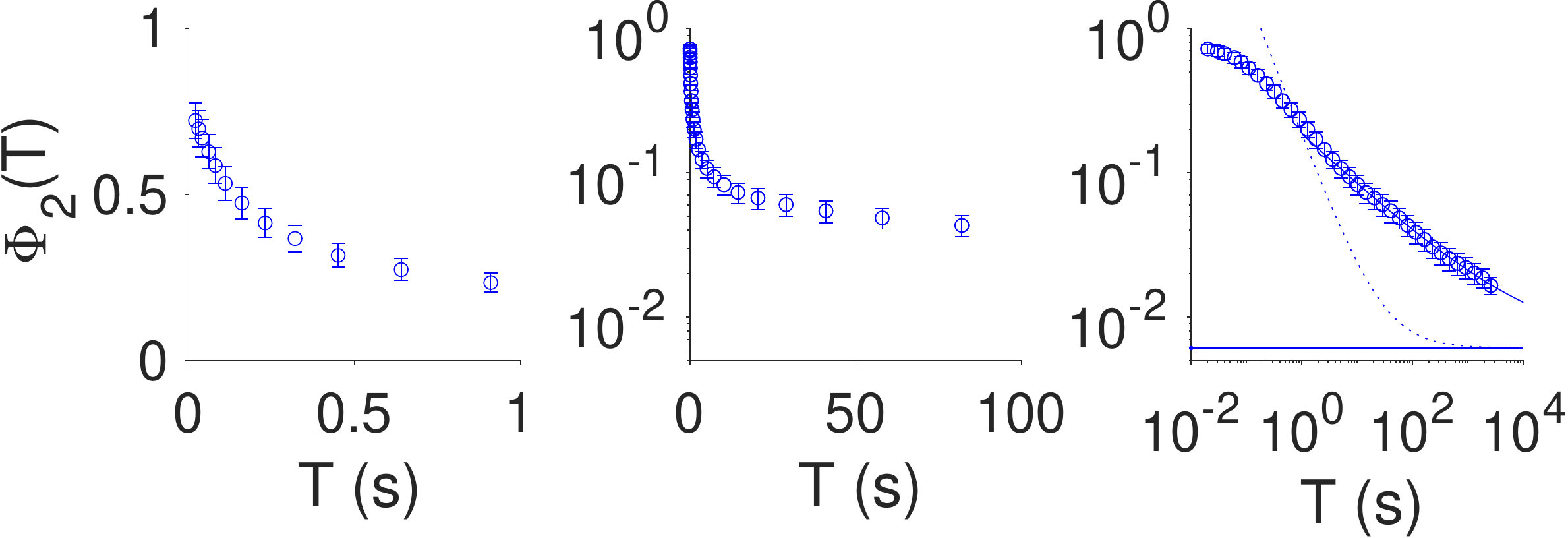}
\caption{Estimates of the summed variance in state probabilities for walking flies as a function of averaging time; raw data from Refs \cite{berman+al_14,berman+al_16}.  Means and standard errors computed from random halves of the data.  (left) Linear plot. (center) Semilog plot. (right) Log--log plot; solid line is Eq (\ref{fit}) for $n=2$ and $T> 7\,{\rm s}$, with parameters in Table \ref{params_table}.
Dashed line shows the estimated plateau $B_2 = \Phi_{2,{\rm ind}}$, and dotted line is a decay $\sim 1/T$ to the same plateau.
\label{fig:Phi2}}
\end{figure*}

We use these ideas to analyze the behavioral state sequences in walking flies described above \cite{berman+al_14,berman+al_16}.
To estimate $\Phi_2(T)$ we follow these steps:
(1) Choose a random window of duration $T$ from the recording of each individual, and estimate the state probabilities. (2) Compute the summed variance of these probabilities across 1000 random halves of the individuals. (3) Use these many random halves to compute a mean and standard error for $\Phi_2(T)$.  Results are in Fig \ref{fig:Phi2}.

On a linear scale (Fig \ref{fig:Phi2}, left),  we see a rapid, sub--second decay of $\Phi_2(T)$.  On a semi--logarithmic plot  (Fig \ref{fig:Phi2}, center) we see  gradual decay out to one minute, but to reveal the full behavior we need a doubly logarithmic plot (Fig \ref{fig:Phi2}, right).  This spans  five decades in time, which is equivalent to measuring correlations between letters from neighboring letters out to the length of a short story.   Beyond $\sim 1\,{\rm s}$ there is no sign of a characteristic time scale, and no clear sign of a plateau at long times, suggesting that genuine individual differences are small.  The decay is much slower than $\Phi_2 \sim 1/T$, suggesting that the system has long--ranged correlations.      Indeed, the data for $T> 7\,{\rm s}$ is an excellent fit to the prediction of Eq (\ref{fit}).

\begin{figure}[b]
\includegraphics[width = \linewidth]{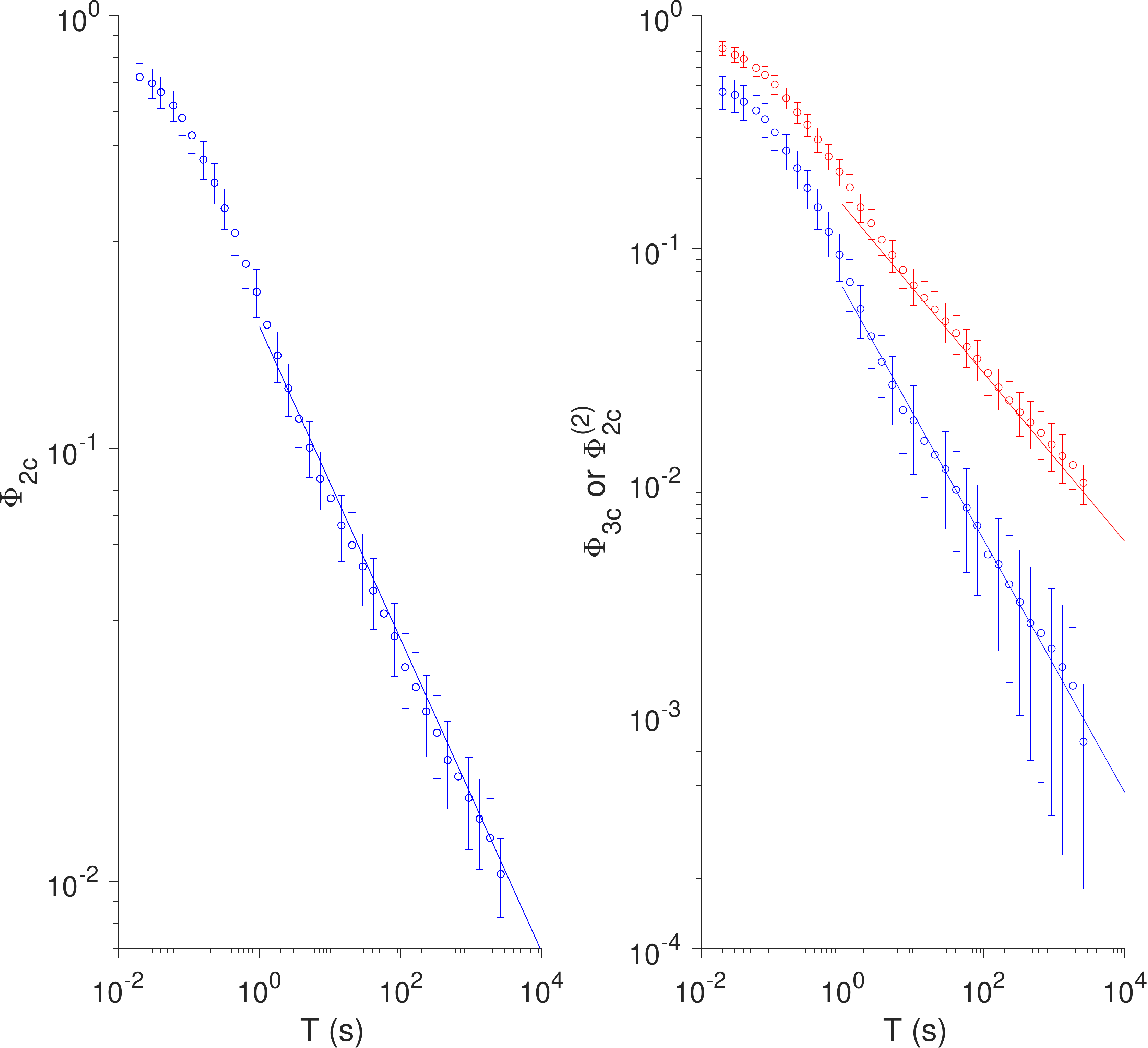}
\caption{Summed variance  in state probabilities as a function of averaging time, with background subtracted. (left) $\Phi_{2c}(T)$. (right)  $\Phi_{3c}(T)$ (blue) and $\Phi_{2c}^{(2)}(T)$ (red).  Error bars include the standard error computed from random halves of the data, as in Fig \ref{fig:Phi2}, and uncertainty in the background.  Lines are best fits as in Fig \ref{fig:Phi2}, parameters from Table \ref{params_table} \label{fig:Phi2c}}
\end{figure}

If our description of $\Phi_2(T)$ is correct, then we can subtract our estimate of the variance across individuals---$B_2$ in Eq (\ref{fit})---to reveal a ``clean'' power--law decay, as shown at left in Fig \ref{fig:Phi2c}.  We see that over three decades in time, all of the data points are within errors of a power--law with exponent $\gamma_2 = 2\Delta$, $\Delta = 0.180 \pm 0.004$.  Since our definition of states uses features on the $\sim 1\,{\rm s}$ time scale \cite{berman+al_14}, and our recordings are $\sim 1\,{\rm h}$ in duration, it is impossible to  see a ``better'' power--law in these data.

In the same way that individual differences can mimic long--ranged correlations, these correlations can mimic individual differences. Naively, if we estimate state probabilities over the hour long experiment, there is a variance across individuals that is $\sim 5\times$ larger than our best estimate of $\Phi_{2,{\rm ind}}$.  It would be tempting to interpret this as biological variability, but this assumes that averaging over one hour is enough to push the statistical fluctuations below the individual variations.  Because of long--ranged correlations, this turns out not to be true.  Our best estimate is that individual differences contribute less than $1\%$ of the total variance in behavior.

Following the discussion above, we can do the same analysis for $\Phi_3(T)$, with the results shown at right in Fig \ref{fig:Phi2c} (blue), and again the fit to Eq (\ref{fit}) is excellent.  We have tried to analyze $\Phi_4(T)$, but the error bars are too large for this to be informative.   If there is a single underlying scale invariant process, then the exponents for $\Phi_2(T)$ and $\Phi_3(T)$ should be $\gamma_2 = 2\Delta$ and $\gamma_3 = 3\Delta$.  Thus  the different moments  provide independent estimates of the underlying scaling dimension $\Delta$, and  Table \ref{params_table} shows that these estimates agree to the third decimal place.

An important lesson from statistical physics is that  structures on long time or length scales are robust to a range of choices  in defining variables on smaller scales.  To test this idea we redefine the ``state'' of the system to be the combined states at two successive moments in time.  Now there are $N_s^2 \sim 15000$ possible states, of which $\sim 1200$ occur in our sample of $2.1\times 10^7$ frames.   We can estimate the probability for each of these states, as before, from data in windows of duration $T$, and define the variance across individuals for these  ``2--frame states,'' $\Phi_2^{(2)}(T)$.  We expect that
\begin{equation}
\langle\Phi_2^{(2)}(T  )\rangle \sim   A_2^{(2)} \left({{\Delta t}\over T}\right)^{2\Delta} +B_{2}^{(2)},
\label{phi22}
\end{equation}
where $\Delta$ is the same scaling dimension as in $\Phi_2 (T)$ and $\Phi_3(T)$, and at right in Fig  \ref{fig:Phi2c} (red) this is confirmed.

In summary, behavioral correlations exhibit precise power--law scaling over three decades in time.  Measurements of three different correlation functions are consistent with  a single underlying scaling field, and estimates of the scaling dimension all agree within the $2-3\%$ experimental errors (Table \ref{params_table}).

\begin{table}[b]
\begin{tabular}{||c|c|c|c||}
\hline
  & $\Phi_2$ & $\Phi_3$ & $\Phi_2^{(2)}$ \\  \hline
 $\Delta$ & $0.180\pm 0.004$ & $0.181\pm 0.005$ & $0.179\pm 0.004$ \\  \hline
 $A$ & $0.936\pm 0.054$ & $0.807\pm    0.047$ & $0.839\pm    0.048$ \\\hline
 $B$ & $0.0060\pm 0.0003$ &$0.0010\pm    0.0001$ & $0.0050  \pm   0.0003$\\\hline
 \end{tabular}
 \caption{Parameter estimates for different measures of variation across individuals.  The underlying scaling dimensions $\Delta$ should agree, while amplitudes $A$ and backgrounds $B$ differ.  Errors  from Monte Carlo sampling of the posterior distribution $\propto \exp(-\chi^2/2)$, with the variance over  random halves of the data used to compute $\chi^2$.  \label{params_table}}
\end{table}

Scale invariance in living systems has a long and controversial history.  Maybe the first example is Zipf's law \cite{zipf_49},  a kind of scale invariance in our vocabulary.  This scaling can be explained away by models of random texts  \cite{miller+chomsky_63,li_92}, but these models  fail to capture other salient features.  Most obviously, not all strings of letters form words \cite{stephens+bialek_10}.  More subtly,  letters  in different words are correlated, and this correlation can be detected over many tens of letters even in modest data sets  \cite{newman+gerstman_52,ebeling+poschel_94}; early evidence  was  hidden in Shannon's 1951 experiments on the prediction of text by human subjects \cite{hilberg_90,shannon_51}.  These data are consistent with the correlations  decaying as a power of the separation between letters, which is a different kind of scale invariance, more closely related to what we have discussed here.  While the scaling vs random text argument reappears periodically, we know of no simple model that captures both the scale invariance of vocabulary and the scale invariance of correlations.  

Temporal correlations that decay as a power law are mathematically equivalent to a mixture of many independent processes happening on different time scales.  These could be trivially independent, as with the electron trapping processes that generate $1/f$ noise in metals \cite{dutta+horn_81}, or they could be the ``normal modes'' of an interacting network so that the spectrum of time scales is an emergent property of the network dynamics \cite{seung_96}.  But true scale invariance is more than a single power--law decay.  In particular, the consistency of scaling dimensions in three different correlation measures is much more difficult to account for with simple mixtures of time scales.

We are surprised by the precision with which scaling behavior emerges from the data.  We have been able to see clean power--law behavior across three decades, from seconds to one hour; errors on exponents are in the third decimal place; and different correlation functions are described by exponents that are consistent with one another, within these small errors, on the hypothesis that there is a single scale invariant process controlling behavior.   Still, we are doing this analysis on data which now are nearly a decade old \cite{berman+al_14}. The next generation of experiments will make continuous measurements of behavior at the same resolution for many weeks rather than for one hour \cite{mckenzie-smith+al_23}.  While it will be necessary to separate circadian and secular variations, these experiments have the potential to test for scaling over six decades, and to reduce errors in the seconds to hours range by more than an order of magnitude, getting us close to the precise tests of scaling in critical phenomena \cite{lipa+al_96}.

\begin{acknowledgments}
We thank GJ Berman, MP Brenner, A Cavagna, I Giardina, O Kimchi,  R Pang,  and FC Pereira for helpful discussions.  This work was supported in part by the National Science Foundation through the Center for the Physics of Biological Function (PHY--1734030),  by the National Institutes of Health through Grants R01EB026943 and R01NS104899, by the John Simon Guggenheim Memorial Foundation, and by the Simons Foundation.
\end{acknowledgments}


\begin{thebibliography}{99}
%
\bibitem{dawkins_76}
R Dawkins, Hierarchical organization: A candidate principle for ethology.  In
{\em Growing Points in Ethology,}  P Bateson and R Hinde, eds,  pp 7--54 (Cambridge University Press, Cambridge,  1976)
%
\bibitem{simon_82}
HA Simon, The architecture of complexity. {\em Proc Am Philos Soc} {\bf 106,} 467--482 (1982).
%
\bibitem{bak_13}
It is hard to find a single origin for this idea in the context of human or animal behavior.  Interest was stimulated by the discovery of self--organized criticality in the mid 1980s, leading to  provocative but perhaps over--ambitious discussions, e.g.
P Bak, {\em How Nature Works} (Springer, New York, 1996).
%
\bibitem{wilson_79}
KG Wilson,  Problems in physics with many scales of length. 
{\em Sci Am} {\bf 241}, 158--179 (1979).
%
\bibitem{cardy_96}
J Cardy, {\em Scaling and Renormalization in Statistical Physics}
(Cambridge University Press, Cambridge, 1996).
%
\bibitem{sethna_06}
J Sethna, {\em Statistical Mechanics: Entropy, Order Parameters and Complexity} (Oxford University Press, 2006).
%
\bibitem{osborne+al_05}
 LC Osborne, SG Lisberger, and W Bialek,    A sensory source for motor variation.  
 {\em Nature} {\bf 437,} 412--416 (2005).
%
\bibitem{stephens+al_08}
GJ Stephens, B Johnson--Kerner, W Bialek, and WS Ryu, Dimensionality and dynamics in the behavior of {\em C. elegans}.    
{\em PLoS Comput Biol} {\bf 4,} e1000028 (2008)
%
 \bibitem{ahamed+al_21}
T Ahamed, AC Costa, and GJ Stephens,  Capturing the continuous complexity of behaviour in {\em C elegans}.  {\em Nat Phys} {\bf  17,} 275--283  (2021).
%
\bibitem{branson+al_09} 
K Branson, AA Robie, J Bender, P Perona, and MH Dickinson,  High--throughput ethomics in large groups of {\em Drosophila}. 
{\em Nat Methods} {\bf 6,} 451--457 (2009).
%
\bibitem{berman+al_14}
GJ Berman,  DM Choi, W Bialek, and JW Shaevitz, Mapping the stereotyped behaviour of freely moving fruit flies.  
{\em J R Soc Interface} {\bf 11,} 20146072 (2014). 
%
\bibitem{berman+al_16}
GJ Berman, W Bialek, and JW Shaevitz,  Predictability and hierarchy in {\em Drosophila} behavior.   
{\em Proc Natl Acad Sci (USA)} {\bf 113,}  11943--11948 (2016).
%
\bibitem{wiltschko+al_15}
AB Wiltschko, MJ Johnson, G Iurilli, RE Peterson, JM Katon, SL Pashkovski, VE Abraira, RP Adams, and SR Datta,   Mapping sub--second structure in mouse behavior.  
{\em Neuron} {\bf 88,} 1--15 (2015).
%
\bibitem{marshall+al_20}
JD Marshall, DE Aldarondo, TW Dunn, WL Wang, GJ Berman, and BP \"Olveczky, Continuous whole--body 3D kinematic recordings across the rodent behavioral repertoire. 
{\em Neuron} {\bf 109,}  420--437 (2021). 
%
\bibitem{reddy+al_20}
G Reddy, L Desban, H Tanaka, J Rouseel, O Mirat, and C Wyart, A lexical approach for identifying behavioural action sequences. {\em PLoS Comput Biol} {\bf 18,} e1009672 (2020).
%
\bibitem{johnson+al_20}
RE Johnson, S Linderman, T Panier, CL Wee, E Song, KJ Herrera, A Miller, and F Engert, Probabilistic models of larval zebrafish behavior reveal structure on many scales. {\em Curr Biol} {\bf 30,} 70--82.e4 (2020).
%
\bibitem{ghosh+rihel_20}
M Ghosh and J Rihel, Hierarchical compression reveals sub--second to day--long structure in larval zebrafish behavior. {\em eNeuro} {\bf 7,} 0408--19 (2020).
%
\bibitem{markowitz+al_13}
JE Markowitz, E Ivie, L Kligler, and TJ Gardner, Long-range order in canary song. {\em PLoS Comput Biol} {\bf 9,} e1003052 (2013).
%
\bibitem{kershenbaum+al_14}
A Kershenbaum, AE Bowles, TM Freeberg, DZ Jin, AR Lameira, and K Bohn, Animal vocal sequences: not the Markov
chains we thought they were. {\em Proc R Soc B} {\bf 281,} 20141370 (2014).
%
\bibitem{kershenbaum+al_16}
A Kershenbaum et al, Acoustic sequences in non-human animals: a tutorial review and prospectus. {\em Biol Rev Camb Philos Soc}  {\bf 91,}  13--52. (2016).
%
\bibitem{mathis+al_18}
A Mathis, P Mamidanna, KM Cury, T Abe, VN Murthy, MW Mathis, and  M Bethge,  DeepLabCut: Markerless pose estimation of user-defined body parts with deep learning.  
{\em Nat Neurosci} {\bf 21,} 1281--1289 (2018).
%
\bibitem{pereira+al_19}
T Pereira, D Aldarondo, L Willmore, M Kislin, SS Wang, M Murthy, and JW Shaevitz,  Fast animal pose estimation using deep neural networks. 
{\em Nat Methods} {\bf 16,} 117--125 (2019).
%
\bibitem{pereira+al_22}
TD Pereira, N Tabris, A Matsliah, DM Turner, J Li, S Ravindranath, ES Papadoyannis, E Normand, DS Deutsch, ZY Wang, GC McKenzie--Smith, CC Mitelut, MD Castro, J D’Uva, M Kislin, DH Sanes, SD Kocher, SS-H Wang, AL Falkner, JW Shaevitz, and M Murthy, SLEAP: A deep learning system for multi-animal pose tracking.  {\em Nat Methods} {\bf 19,} in press (2022).
%
\bibitem{brown+bivort_17}
AEX Brown and B de Bivort,  Ethology as a physical science. 
{\em Nat Phys} {\bf 14,} 653--657 (2017).
%
\bibitem{berman_18}
GJ Berman, Measuring behavior across scales. {\em BMC Biology} {\bf16,}  23 (2018).
%
\bibitem{datta+al_19}
SR Datta, DJ Anderson, K Branson, P Perona, and A Leifer,  Computational neuroethology: A call to action. 
{\em Neuron} {\bf 104,} 11--24 (2019).
%
\bibitem{pereira+al_20}
TD Pereira, JW Shaevitz, and M Murthy,  Quantifying behavior to understand the brain.  
{\em Nat Neurosci} {\bf 23,} 1537--1549 (2020).
%
\bibitem{mathis+mathis_20}
MW Mathis and A Mathis,  Deep learning tools for the measurement of animal behavior in neuroscience.
{\em Current Opin Neuro} {\bf 60,} 1--11 (2020).
%
\bibitem{bialek_20}
W Bialek,  On the dimensionality of behavior. {\em Proc Natl Acad Sci (USA)} {\bf 119,} e2021860119 (2022); arXiv:2008.09574 [q--bio.NC] (2020).
%
\bibitem{alba+al_20}
V Alba, GJ Berman, W Bialek, and JW Shaevitz, Exploring a strongly non--Markovian animal behavior.   arXiv:2012.15681 [q--bio.NC] (2020).
%
\bibitem{cavagna+al_18}
A Cavagna, I Giardina, and TS Grigera, The physics of flocking: Correlation as a compass from experiments to theory. {\em Phys Repts} {\bf 728,} 1--62 (2018).
%
\bibitem{hernandez+al_21}
DG Hern\'andez, C Rivera, J Cande, B Zhou, DL Stern, and GJ Berman, A framework for studying behavioral evolution by reconstructing ancestral repertoires. {\em eLife} {\bf 10, } e61806 (2021).
%
\bibitem{intC0}
The same  issue arises in analyzing the spatial correlations of velocity fluctuations in a flock of birds \cite{cavagna+al_18}. 
%
\bibitem{zipf_49}
GK  Zipf, {\em Human Behavior and the Principle of Least Effort} (Addison-Wesley, Reading, MA, 1949).
%
\bibitem{miller+chomsky_63}
GA Miller and N Chomsky, Finitary models of language users. In {\em Handbook of Mathematical Psychology}, RD Luce et al, eds, pp 419--491 (Wiley, New York, 1963).
%
\bibitem{li_92}
W Li, Random texts exhibit Zipf's--law--like word frequency distribution.  {\em IEEE Trans Inf Theory} {\bf 38,} 1842--1845 (1992).
%
\bibitem{stephens+bialek_10}
GJ Stephens and W Bialek,  Statistical mechanics of letters in words.  {\em Phys Rev E} {\bf 81,} 066119 (2010)
%
\bibitem{newman+gerstman_52}
EB Newman and LJ Gerstman,  A new method for analyzing printed English. {\em  J Exp Psychol} {\bf 44,} 114--125 (1952).
%
\bibitem{ebeling+poschel_94}
W Ebeling and T P\"oschel, Entropy and long--range correlations in literary English.   {\em Europhys Lett} {\bf 26,} 241--246 (1994).
%
\bibitem{shannon_51}
CE Shannon, Prediction and entropy of written English.  {\em Bell Sys Tech J} {\bf 30,} 50--64 (1951).
%
\bibitem{hilberg_90}
W Hilberg,  Der bekannte Grenzwert der redundanzfreien Information in Texten: eine Fehlinterpretation der Shannonschen Experimente?.  {\em Frequenz} {\bf 44,} 243--248 (1990).
%
\bibitem{dutta+horn_81}
P Dutta and PH Horn, Low--frequency fluctuations in solids: $1/f$ noise. {\em Rev Mod Phys} {\bf 53,} 497--516 (1981).
%
\bibitem{seung_96}
For  modes in neural networks, and especially the emergence of modes that can carry correlations over long times, see e.g. HS Seung, How the brain keeps the eyes still.   {\em Proc Natl Acad Sci (USA)} {\bf 93,} 13339--13334 (1996).
%
\bibitem{mckenzie-smith+al_23}
G McKenzie--Smith, SW Wolf, JF Ayroles, and JW Shaevitz, Long term recordings of {\em Drosophila melanogaster} behavior at high temporal resolution. {\em APS March Meeting} B08.00011 (2023).
%
\bibitem{lipa+al_96}
JA Lipa, DR Swanson, JA Nissen, TCP Chui, and US Israelsson, Heat capacity and thermal relaxation of bulk helium very near the lambda point. {\em Phys Rev Lett} {\bf 76,} 944--947 (1996).  
%
\end{thebibliography}
\end{document}